\documentstyle [12pt] {article}
\begin {document}

\large
\parindent 0 cm
\begin {center}
{\bf Fractal Dimension of Backbone of Eden Trees} \\
\vskip 1.0 cm
\normalsize
S. S. Manna$^1$ and D. Dhar$^2$
\vskip 1.0 cm
     $^1$Department of Physics, Indian Institute of Technology \\
     Powai, Mumbai 400076, India \\
     e-mail : manna@niharika.phy.iitb.ernet.in
\vskip 0.2 cm
     $^2$Theoretical Physics Group, Tata Institute of Fundamental
     Research \\ Homi Bhabha Road, Mumbai 400005, India \\
     e-mail : ddhar@theory.tifr.res.in
     
\vskip 1.0 cm
{\underline {Abstract}}

\vskip 0.5 cm
\end {center}

\normalsize
     
\parindent = 1 cm
     We relate the fractal dimension of the backbone, and the
     spectral dimension of Eden trees to the dynamical exponent
     $z$. In two dimensions, it gives fractal dimension of
     backbone equal to $4/3$ and spectral dimension of
     trees equal to $5/4$. In three dimensions, it provides us a new
     way to estimate $z$ numerically. We get $z=1.617 \pm 0.004$.
\vskip 1.0 cm
     Dense branching patterns are found in many different
     physical situations in nature e.g., coral reefs, river
     networks, collapsed phase of branched polymers, very slowly
     evaporated films of sugar dissolved in water [1-3]. In all
     these systems, the Hausdorff dimension of the structure is
     equal to that of the embedding space but
     the detailed structure is different depending on the
     different physical processes involved.

 The Eden model has been studied a lot in the
     last decade, mainly for the surface properties [4]. 
Eden trees [5,6] are simple theoretical model of dense branching
     structures.
     In [5], it was argued that
     classical diffusion on Eden trees is anomalous because of
     trapping in dead end branches, and the root mean square
     deviation of a random walker on the tree increases with
     time as $t^x$, where the exponent $x$ does not satisfy the
     usual relation $x = \tilde d/2\bar d$, where $\tilde d$ is
     the spectral dimension, and $\bar d$ is the (Hausdorff)
     fractal dimension of the lattice. Dhar and
     Ramaswamy expressed the exponents $x$ and $\tilde d$ in
     terms of an exponent $\theta$ related to the fractal
     dimension of the backbone of the trees [5]. 
Using a different method of analysis, and
     somewhat larger simulations, Nakanishi and Herrmann [6]
     also calculated these exponents. 
However, these exponents
     have not been determined analytically so far.

In this Rapid Communication, we show that the backbone exponent of the
     Eden trees can be related to the dynamical exponent of KPZ
     model, and show that in two dimensions,$\theta = 1/3$, $\tilde d = 5/4$ 
and $x =3/8$. 
In three dimensions, our method gives us a new way to determine
     the dynamical exponent numerically. The numerical
     determined value is $ z = 1.617 \pm 0.004 $.

We shall consider {\it {Bond}}-Eden Trees (BET) in this report. The
model is defined in terms of the spreading of an infectious disease along
the bonds of a $d$-dimensional lattice. Each site may be in one of two
states : healthy or infected. At time t=0, a fixed set of `seed' sites are
infected. This seed set is a single site in the so-called point seed
geometry, and a $(d-1)$ dimensional hyper-plane in layer seed geometry. At
each time-step, we select at random one site from the set of healthy sites
having at least one infected neighbour. This site then becomes infected,
and once infected, a site never recovers. We connect this site to existing
infected cluster by the bond connecting it to its first-infected neighbour.
Each new infected site adds exactly 1 bond to the cluster. The resulting
cluster so generated is called a BET. 
 
However, the bulk of the Bond Eden trees is a {\it {spanning tree}}, and
     has a complex internal structure. This may be quantified in terms
     of the fractal dimension of the chemical paths along the
     tree, distribution of branch sizes, spectral dimension of
     the tree, structure of its backbone etc. We shall see below
     that all these measures can be determined in terms of
     a single critical exponent, which we choose to be the dynamical
     critical exponent $z$.

There is a unique path which connects any two sites of a tree.
     We define the backbone of the tree corresponding to radius $R$ 
as the set of all sites
     which lie on a path connecting one of the seed sites to any
     of the sites of the tree at a distance $R$ from it. Fig. 1 
shows the backbone of a two
     dimensional Eden tree grown from a single seed for two
     values  $R = 80 $ and $130$
     lattice spacings. As the cluster grows, 
     some of the growing
     branches became dead ends, and are removed from the
     backbone. While the part of the backbone near the surface
     changes quite fast, structure of the backbone deep inside
     the cluster does not change much with time, and gets {\it
     {frozen in}}.

From  Fig. 1, it is easy to see that the backbone of Eden
     clusters has a branching structure, where each branch is
     directed radially outwards. A branch of the backbone of
     length $R$ has transverse fluctuations of order $R^{1/z}$,
     where clearly $z \ge 1$ and is called the dynamical exponent.
This exponent is inverse of the exponent specifying the transverse
fluctuations of a directed polymer in a random medium [7].

     We now argue that the backbone has a fractal dimension
     $d_B$ given by 

\begin {equation}
     d_B = 1 + (d-1)(1-1/z)
\end {equation}
Consider the backbone of the cluster when its radius is $R$, and
     at a later stage when its radius is $R+h, h << R$. Only a
     small fraction of the perimeter sites at radius $R$ remain
     part of the backbone at the later stage. Each such site
     gives rise to a cluster of active sites of transverse size
     $h^{1/z}$. This implies that the density of backbone sites
     at radius $R$, when cluster has grown to size $(R+h)$
     vanishes as $h^{-(d-1)/z}$ for large $h << R$. Once $h$ is
     of order $R$, it gets frozen to the value $R^{-(d-1)/z}$,
     and does not change further. Thus the number of sites in
     the frozen backbone upto radius $R$ varies as $R^{d_B}$,
     where $d_B$ is given by Eq.(1). 

The spectral dimension of the Eden trees is clearly a bulk
     property and can also be expressed in terms of $z$.
     Consider a spring network on the tree, such that at each
     bond of the tree, there is a spring of spring constant $\kappa$,
     and a mass $m$ is attached to each site of the tree. Let
     $F(\omega^2)$ be the fractional number of eigenvalues with
     frequency less than $\omega$. Then if $F(\omega^2) \sim
     \omega ^{\tilde d}$ for small $\omega$, $\tilde d$ is the
     spectral dimension of the tree.

To determine $\tilde d$, we use a decimation argument analogous
     to that in [8]. Define the `burning time' at a site on the
     tree $T$ as the length of the longest directed path from the site to
     a leaf site of the tree. We construct a decimated tree $T'$, 
consisting of
     only those sites whose burning time is an exact multiple of
     an integer scale factor $b$. The resulting tree $T'$ is not
     related to $T$ by a simple scale change (for some figures
     of decimated trees see [9]). However, small parts of $T$
     are related to corresponding parts of $T'$ by local
     self-affine transformation, where radial distances are
     scaled by a factor $b$, and transverse distances
     by a factor $b^{1/z}$. The spring constants
     renormalize by $\kappa' = \kappa/b$, and masses 
     according to the formula 
     $M' = b^{1+(d-1)/z}M$. Thus we get 

\begin{equation}
F(\kappa'/M') = b^{-1-(d-1)/z}F(\kappa/M).
\end{equation}
     
which implies that 

\begin {equation}
\tilde d = 2. \frac {z+d-1}{2z+d-1}. 
\end{equation}
     
     For $d=2$, the dynamical exponent takes the well known KPZ
     value 3/2. Correspondingly, we get $d_B = 4/3$ and $\tilde d
     = 5/4$. This value is in good agreement with the earlier
     numerical estimates of [5] and [6]. 
Dhar and Ramaswamy found that for
     $d$ = 2, $\tilde d = 1.22 \pm 0.04, x = 0.42 \pm
     0.04$, and for $d = 3, \tilde d = 1.30 \pm 0.12, x = 0.44
     \pm 0.04$ [5]. 
Nakanishi and Herrmann estimated $ d = 2, 
\tilde d = 1.22 \pm 0.02$ and $x
     = 0.39 \pm 0.02$ where as for $ d = 3, \tilde d = 1.32
     \pm 0.02$ and $x = 0.30 \pm 0.02$ [6]. 
As $d \rightarrow
     \infty, \tilde d$ tends to 2, which again agrees with the
     exact result [10].

For the root mean square displacement of order $L$ the average
     number of branch site per backbone site visited is of order
     $L^{(d-1)/z}$. Hence the diffusion constant is of order
     $L^{-(d-1)/z}$ and time $T$ scales as $L^{2+(d-1)/z}$, so
     we get $ 1/x = 2 + (d-1)/z$. We note that $x \ne \tilde d /
     2 \bar d$, where $\bar d$ is the fractal dimension of the
     substrate. However, the equality sign holds if $\bar d$
     is replaced by $d_m = 1 + (d-1)/z$, which is the effective
     mass dimension of the graph defined by the relation that
     the number of distinct sites within a distance $r$ along
     the tree from a randomly chosen site (not near origin)
     varies as $r^{d_m}$ for large $r << $ size of cluster $R$.

We have checked these predictions against numerical simulations.
     Figure 2 shows a plot of the average number of sites in the
     backbone $<M_b>$ of a 2-dimensional Bond Eden tree within a (Euclidean)
     distance $R$ from the origin multiplied by
     a factor $R^{-4/3}$ versus $R^{-1/3}$. We averaged over
     12 million independently generated configurations for 
$R= 16$ decreased to 12000 configurations for $R$ = 1024 
on the square lattice. 
We see a reasonably good fitting of a straight line
in fair agreement with the theoretical
prediction of $d_B$ =4/3 which also indicates that the correction
to scaling is likely to be $R^{-1/3}$. We estimate the error 
in the quoted values of both the fractal dimension 
and the correction to scaling exponent to be about 0.03.

In Fig. 3 we show
     the results of simulations of Eden trees on a triangular
     lattice grown from a line seed on a $6000 \times 10000$
     lattice, averaged over 2500 configurations. We have plotted
     $ h^{2/3} N_h$ versus $h$, where $N_h$ is the fractional number of
     distinct trees which survive up to height $h$. From the
     scaling hypothesis $N_h \sim h^{-\alpha}$,
     where $\alpha = (d-1)/z$  In this case $\alpha = 2/3$, and hence the
     graph should be a horizontal line. For large $h$ we see that
     this expectation is very well satisfied, and the
     numerically determined value of $z$ from this plot gives
     $z = 3/2 \pm 0.002$.

The main advantage of our simulation over earlier simulations of
     various versions of Eden model, e.g., models A, B and C of
     Jullien and Botet [12] is that here we are in effect
     studying bulk quantities (The quantity $N_h$ can be
     related to the probability that randomly chosen site in the
     bulk of the
     tree has at least one descendant left after $h$ more
     generations). A related quantity, the distribution of
     branch sizes which are disconnected on removing a randomly
     chosen bond from the tree has been studied in [9].

Encouraged by the good convergence of simulation results in our
     model to the asymptotic values in 2-dimensions, we extended
     our studies to 2+1 dimensions. In Fig. 4, we plot the
     results of simulation of Eden growth in 2+1 dimension in
     the layer geometry. We used as $180 \times 180 \times 500$
     simple cubic lattice, and averaged over $10500$
     configurations. Fig. 4 shows
     $ h^{5/4} N_h$ versus $h$ on a log-log plot.
     We estimate the slope of the curve in this plot to be
     $0.013 \pm .003$. This implies that 
     $\alpha = 5/4 -0.013 \pm 0.003$,
     which corresponds to $z = 1.617 \pm .004$. The value
     $\alpha$ = 5/4 corresponds to the Kim-Kosterlitz 
     conjecture [13], which is clearly ruled out by our data. Our
     values are in good agreement with the current best
     numerical estimate of $\alpha = 1.240 \pm 0.001$ 
     by Forrest and Tang [14] and Ala-Nissila and Venalainen's result
     of $\alpha = 1.240 \pm 0.002$ [15].

     The extension to higher dimensions seems
     possible, but would require higher computational power
     than available to us at present.

We thank Drs. T. Halpin-Healy and L. H. Tang for providing some
     useful references.
\vfill
\eject

\parindent= 0 cm
{\underline {\bf Figure Captions }}
\vskip 1.0 cm

Figure 1 : Backbone of a Bond Eden Tree generated on the square
lattice from a point seed at two stages of growth of radii 80 and
130 lattice constants. 
\vskip 0.5 cm
Figure 2 : Variation of average mass of the backbone $<M_b>$ of 
Bond Eden Trees of radius $R$. The fractal dimension
of the backbone is $\approx 4/3$, in complete agreement with
our theoretical prediction and the correction to scaling decreases as  
$R^{-1/3}$.
\vskip 0.5 cm
Figure 3 : Plot of $h^{2/3}N_h$ versus $h$ for Bond Eden Trees grown 
on a base line of length 6000. Here $N_h$ is the fractional number of 
distinct trees which survive upto height $h$ . The average slope of this plot
for  $h$ between 1000 and  10000 is at most $\pm 0.003$.
\vskip 0.5 cm
Figure 4  :Plot of $h^{5/4}N_h$ versus $h$ for Bond Eden Trees  grown
on a square base of size $ 180 \times 180$. 
The average slope
between $h$ = 100 and 500 is 0.013. A straight line of slope 0.013 is
plotted as a guide to the eye. 
\vfill
\eject

\parindent= 0 cm
{\underline {\bf References }}
\vskip 1.0 cm
     
[1.] S. S. Manna and B. Subramanian, Phys. Rev. Lett. {\bf 76}
     (1996) 3460;
     A. Rinaldo, I. Rodriguez-Iturbe, R. Rigon, E.
     Ijjasz-Vasquez and R. L. Bras, Phys. Rev. Lett. {\bf 70},
     822 (1993);
     A .Giacometti, A. Maritan, and J. R. Banavar, Phys. Rev.
     Lett. {\bf 75}, 577 (1995).

[2.] B. Derrida and H. J. Herrmann, J. Phys. (Paris) 
     {\bf 44} (1983) 1365. 

[3.] A. S. Paranjape, Phys. Lett. A {\bf 176} (1993) 349. 

[4.] For a review see P. Meakin, Phys. Rep. {\bf 235}
     (1993) 189; T. Halpin-Healy and Y. C. Zhang, Phys. Rep.
     {\bf 254} (1995) 215.

[5.] D. Dhar and R. Ramaswamy, Phys. Rev. Lett. {\bf 54}, (1985) 1346.

[6.] H. Nakanishi and H. J. Herrmann, J. Phys. A {\bf 26}
     (1993) 4513. 

[7.] S. Roux, A. Hansen and E. L. Hinrichsen, J. Phys. A {\bf 24}, (1991)
     L295; M. Cieplak, A. Maritan and J. R. Banavar, Phys. Rev. Lett.
     {\bf 76}, 3754 (1996).

[8.] S. S. Manna, D. Dhar and S. N. Majumdar, Phys. Rev. A
     {\bf 46} (1992) R4471. 

[9.] P. Meakin, Physica Scripta {\bf 45} (1992) 733. 

[10.] D. Dhar, J. Phys. A {\bf 18} (1985) L713.

[11.] J. Krug and P. Meakin, Phys. Rev. A {\bf 40} (1989) 2064. 

[12.] R. Jullien and R. Botet J. Phys. A {\bf 18} (1985)
     2279. 

[13.] J. M. Kim and J. M. Kosterlitz, Phys. Rev. Lett. {\bf
     62} (1989) 2289. 

[14.] B. M. Forrest and L. H. Tang, Phys. Rev. Lett {\bf 64}
     (1990) 1405. 

[15.] T. Ala-Nissila and O. Venalainen, J. Stat. Phys. {\bf 76}
(1994) 1083.

\end {document}